\documentclass[10pt, b5paper]{article}
\usepackage{amsmath ,amsthm ,amssymb}
\usepackage{setspace, longtable, graphicx, hyphenat, hyperref, fancyhdr, ifthen, everypage, enumitem, amsmath, setspace}
\usepackage{graphicx}
\usepackage[margin=2.5cm, width=12.5cm, height=19.0cm]{geometry}
\usepackage{xcolor}
\usepackage{relsize}
\usepackage{bbm}
\usepackage[utf8]{inputenc}
\usepackage{libertine}
\usepackage[libertine]{newtxmath}
\usepackage{stmaryrd} 
\usepackage{tikz-cd}
\usepackage{tikz}
\usepackage{braket}
\usepackage{cleveref}
\usetikzlibrary{matrix}
\usepackage{bbold}
 \usepackage{hyperref} 
\hypersetup{
	colorlinks=true,
	linkcolor=blue,
	filecolor=blue,
	citecolor=cyan,      
	urlcolor=cyan,
}

\title{Particles in the superworldline and BRST}
\author{Eugenia Boffo \\ \small Faculty of Mathematics and Physics, Mathematical Institute, \\ \small Charles University Prague, Sokolovsk\'{a} 49/83, 186 75 Prague.\\ \small {boffo@karlin.mff.cuni.cz}}
\date{}

\begin{document}
	\maketitle
	
	\begin{abstract}
		In this short note we discuss $N$-supersymmetric worldlines of relativistic massless particles and review the known result that physical spin-$N/2$ fields are in the first BRST cohomology group. For $N=1,2,4$, emphasis is given to particular deformations of the BRST differential, that implement either a covariant derivative for a gauge theory or a metric connection in the target space seen by the particle. In the end, we comment about the possibility of incorporating Ramond-Ramond fluxes in the background.  
	\end{abstract}
	
	
	\section{Introduction}
	
	In string theory, scattering amplitudes are calculated by inserting on the worldsheet, describing the propagation of a free string, the vertices for external states. Although such a procedure is common in the context of string theory, it is not exclusive of Riemann surfaces and can be exported to particles' worldlines. The advantages of expressing a physical model on a worldsheet or worldline are not limited to the calculation of scattering amplitudes. Covariant quantization is also easy to perform in this context. The Becchi-Rouet-Stora-Tyutin (BRST) quantization \cite{Becchi:1975nq}, or the Batalin-Vilkovisky (BV) quantization \cite{Batalin:1983ggl} in more convoluted cases, are powerful methods that allow to perturbatively compute the path integrals of the worldsheet/worldline theory. BRST or BV supersede canonical quantization with constraints if there is degeneracy due to 
	\emph{gauge symmetries}. Often the background geometry is fixed from the outset and it is taken to be flat, but thanks to the cohomological nature of the aforementioned BRST and BV methods, deformations of the background can be handled too. These are taken into account in the differential for the complex. In particular, connection $1$-forms with values in a Lie algebra are easy to implement. The present article aims at exploring a few instances of curved backgrounds for $N$-superworldines, with $N=1,2,4$.
	In fact, particles constitute an excellent playground. The smaller number of invariances that they enjoy, compared to strings, facilitate the investigation while at the same time giving good insights for the string case.
	
	This note is organized as follows: After presenting the so called \emph{$N$-spinning particle} model in \cref{2}, i.e. an $N$-supersymmetric worldline, in \cref{3} we will review the background deformations studied in the literature for $N=1,2,4$, and conclude in \cref{4} by hinting at a strategy which implements R-R fluxes in the background. We will keep the exposition quite basic and mainly address an audience of theoretical physicists, at the same time stressing the underlying algebraic and geometric aspects.  
	
	\section{Particles in the $N$-superworldline}
	\label{2}

	In the `70es Brink, Di Vecchia and Howe \cite{Brink:1976uf} first showed that in the worldline formalism, a massless particle with $N=1$ charge of supersymmetry describes a Dirac spinor, satisfying the Dirac equation, upon first quantization. This observation was later generalized to more values of $N \in \mathbb{N}$ and quantum equivalence with the spin-$N/2$ particle proven. The interested reader can look at the review article \cite{Corradini:2015tik} and references within.
	
	Let us be pedestrian and describe the field content and the Lagrangian theory for generic $N$ in the positive integers. The fields in the model consists of maps from the ($N$-)superworldline to target space, which might have either even or odd parity, according to the $\mathbb{Z}_2$ grading. Being functions on a supermanifold, in fact, they should be thought as sections of a vector bundle over a regular base manifold $M_0$. The structure sheaf of a $N$-supermanifold, if $U_0 \subset M_0$ is an open contractible subset, consists of smooth functions of $U_0$ in tensor product with the symmetric algebra of a graded vector space $V$:
	\[
	C^\infty(U_0) \otimes \odot^\bullet V.
	\]
	Thus, for the $N$-superworldline with Grassmann even and odd coordinates respectively $(\tau \, , \, \theta_\mathrm{k})_{ \mathrm{k} = 1, \dots N}$, we are interested in maps of total degree $0$ and hence parity even. Moreover in the present note we will not need to go beyond linear order in $\theta_\mathrm{k}$:
	\begin{equation}
		X^\mu(\tau) + \theta_\mathrm{k}\Psi^\mu_{\mathrm{k}}(\tau), \qquad 		X^{\mu}(\tau), \, \Psi^{\mu}_{\mathrm{k}} (\tau) : \left(\mathrm{I}, \mathrm{d}s^2\right) \mapsto \left(M, g\right),
	\end{equation}
	where $\mathrm{I} \subset \mathbb{R}\subset \mathbb{R}^{1\vert N}$ is a real interval of the line embedded in the ($N$-)superline, and $\mathrm{d}s^2$ its metric. $M$ is a $4$-dimensional metric manifold with pseudo-Riemannian metric $g$. The $N$ fields $\Psi_\mathrm{k}$ are parity odd and thus satisfy, at the same instant and for fixed k, the graded commutativity relation: \begin{equation*}\Psi^{\mu}_{\mathrm{k}}(\tau) \Psi^{\nu}_{\mathrm{k}}(\tau) = - \Psi^{\nu}_{\mathrm{k}}(\tau) \Psi^{\mu}_{\mathrm{k}}(\tau). \end{equation*}
	
	The smooth maps can be organized, following the principle of invariance, in a Lagrangian and subsequently an action functional. We require invariance under local $\mathrm{Diff}(\mathbb{R})$ and local supersymmetry. To make this happen, additional fields must be introduced: $e(\tau)$, the einbein of $\mathrm{I}$, that takes into account the freedom in the parametrization of the line, and $\chi_{\mathrm{j}}(\tau)$, which are $j$ Grassmann odd variables for local supersymmetries. Furthermore, we will present the action in the so-called \emph{first order formalism}, whose name is reminiscent of the fact that it yields first order differential equations. Namely there is another function $P_\mu$ of even Grassmann degree, so that $(X^\mu, P_\nu)$ are coordinates for $T^*M$.
	
	For simplicity, take $N=1$ and the Lorentzian metric $g = \mathrm{d}x^{\mu} \eta_{\mu \nu} \mathrm{d}x^{\nu}$. Then, after integration over the odd coordinates, the action reads:
	\begin{equation}
		\mathrm{S} = \int_{\mathrm{I}} \, \mathrm{d} \tau \, P_{\mu} \dot{X}^{\mu} + \mathrm{i} \Psi^{\mu} \dot{\Psi}_{\mu} - \frac{e}{2} P^{2} - \mathrm{i} \chi \Psi^{\mu} P_{\mu}. \label{sp-part}
	\end{equation}
	Under infinitesimal diffeomorphisms of the line $\mathrm{I}$, with vector field $Y\partial_\tau$, the fields transform as:
	\begin{align}
		\delta X^\mu = Y \dot{X}^\mu, & \, & \delta e = Y \dot e + \dot Y e, \notag \\
		\delta \Psi^\nu = Y \dot{\Psi}^\nu, & \, & \delta \chi = Y \dot \chi + \dot{Y}\chi. \label{diffeo}
	\end{align}
	Instead the infinitesimal variations $\delta_\epsilon$ of the fields under supersymmetry, generated by an odd parameter $\epsilon(\tau)$, look like:
	\begin{align}
		\delta_\epsilon X^\mu = \mathrm{i} \epsilon \Psi^\mu , & \, & \delta_\epsilon e = 2 \mathrm{i} \chi \epsilon, \notag \\
		\delta_\epsilon \Psi^\mu = - \epsilon P^\mu, &\,& \delta_\epsilon \chi = \dot{\epsilon}. \label{susy}
	\end{align}
	Invariance of \eqref{sp-part} under the symmetries given above is easily checked. 
	
	The classical dynamics in gauge fixed form, $e=1$ and $\chi =0$, corresponds to:
	\[
	\dot{P}^{\mu}(\tau) = 0, \quad \dot{\Psi}^{\mu}(\tau) = 0,
	\] 
	i.e. free massless particle and constant, parity odd variables.
	
	Concerning the first quantization of the system, one can resort to canonical constrained quantization. The orthosymplectic algebra:
	\begin{equation}
		[X^\mu, P_\nu] = \mathrm{i} \delta^\mu_\nu , \qquad \{\Psi^\mu, \Psi^\nu\} = \eta^{\mu\nu},
	\end{equation}
	can be represented on the space of spinors. A state in the (infinite-dimensional) Hilbert space is generated from the highest weight vector $\ket{0}$ as
	\begin{equation}
		\ket{\rho} = e^{-\mathrm{i} P\cdot X} u(P) \ket{0}, \label{Dirac}
	\end{equation}
	where $u(P)$ is a spinor. Then the (super)algebra of the constraints $H := P^2$ and $q_0 := \Psi^\mu P_\mu$,
	\begin{equation}
		\{q_0, q_0\} =H, \qquad [q_0, H ] =0,
	\end{equation}
	imposes the conditions that $\ket{\rho}$ satisfies $\Psi^\mu P_\mu \ket{\rho} =0$, which is the Dirac equation, as well as the massless Klein-Gordon equation $ P^2 \ket{\rho} = 0$.

	\section{BRST cohomology}
	\label{3}
	
	For the classical particle model presented in \eqref{sp-part}, old constrained (Dirac) quantization is sufficient to disclose its quantum mechanical aspects, especially the representation on spinors. Nevertheless the technique of BRST quantization \cite{Figueroa} can also be applied. It was devised in order to obtain an explicitly covariant quantization of regular gauge theories, whereas here the algebra of symmetries is $\mathbb{Z}_2$-graded, but the standard formulation is extended naturally without any effort to graded/super Lie algebras. We already pointed out in the introduction that one of the strengths of BRST resides in the possibility of investigating various different backgrounds. Before digging into this, a quick intro is due. As it stands, however, BRST cohomology is a vast topic on its own and a precise introduction is an enormous task, beyond the scopes of the present note. Let us anyway try to lay down some of the fundamentals: one wants to have a double complex $C^{p,q}$ made up with a projective resolution of the $\mathfrak{g}$-module $C^\infty(M)$ for the (super)Lie algebra $\mathfrak{g}$, where each term is in tensor product with $\Lambda^p\mathfrak{g}^*[1]$.\footnote{The number in square brackets is a shift by $1$ of the vector space, so that for example at $p=0$ we have functionals $\mathfrak{g}[1] \mapsto \mathbb{R}$. Effectively it means that if $\mathfrak{g}$ is a regular (ungraded) algebra, now the basis has degree $-1$ and thus the coordinate functions are of degree $+1$ and their parity is hence odd.} The total complex is $\mathcal{C}^\bullet$, sum over all $p$ and $q$ of $ C^{p,q} = \Lambda^p\mathfrak{g}^*[1] \otimes \Lambda^q \mathfrak{g}[1] \otimes C^\infty(M)$, with a total differential $D$ that sends $\mathcal{C}^{p,q}$ with fixed $p-q$ to those with $p-q+1$ (though it can increase $p$ by 1 and decrease $q$ by $-1$). By construction, there is an isomorphism of the degree $0$ cohomology groups:
	\[ H^0(\mathcal{C}^\bullet, D) \cong H^0(\mathfrak{g}, C^\infty(M^0)) \cong C^\infty(\tilde{M}), \]
	where $C^\infty(M^0)$ are the functions of $M$ modulo those that vanish on a closed embedded submanifold $M^0$, and $\tilde M = M // G$ is the symplectic reduction, for $G$ the Lie group integrating the Lie algebra $\mathfrak{g}$. Hence studying the cohomology of the double complex is equivalent to construct the smooth functions on $\tilde M$. Furthermore, $D$ is given by the graded Poisson bracket of an element $Q \in \mathcal{C}^1$, thus having total "ghost number"\footnote{In Physics' parlance, the ghost number is $- 1$ for basis elements of $\mathfrak{g}$ and $+1$ for its dual.} equal to 1, $D = \{Q, -\}$. $D^2=0$ follows from $\{Q,Q\}= 2 Q^{2} = 0$, and can be seen as an "enhanced" Chevalley-Eilenberg differential. Then $Q$ itself serves as the differential operator (derivation for a graded Poisson algebra).
	
	For our purposes, we will make use of the cohomological nature and especially of the differential in order to consider curved target spaces. Furthermore, since we are in the conditions where quantization (done by replacing the Poisson algebra of functions with the commutator algebra of observables, which acts on a Hilbert space) "commutes" with symplectic reduction \cite{LOSEV20121216}, we will be talking about this quantized setting.
	
	\bigskip
	
	Practically, in the present case where the super Lie algebra is that of superdiffeomorphisms of $\mathbb{R}^{1 \vert N}$ (see again \eqref{diffeo} and \eqref{susy} for $N=1$), we need to introduce a pair of ghost-antighost fields for each conserved charge, namely the total energy $P^2$ and the $N$-supercharges $\Psi^{\mu}_{i} P_{\mu}$. The parity of the new variables will be opposite to that of the charges, which is odd for the latter ones. Eventually we should enlarge the orthosymplectic algebra with the set $c,b,\beta_k, \gamma_j$ and the relations:
	\begin{equation}
		\{\overset{+1}{c^{}}, \overset{-1}{b}\} = 1, \quad [\overset{+1}{\gamma_{j}^{}},\overset{-1}{\beta_{k}}] = \delta_{jk}.
		\label{alg2}
	\end{equation}
	On top of each generator the respective ghost number is displayed. Together with the Clifford algebra of $\Psi_{k}$ and the canonical Weyl algebra of positions $X^{\mu}$ and momenta $P_{\nu}$ of $T^{*}M$, \eqref{alg2} gives the BRST algebra relevant for this case.
	
	Let us discuss $Q$ for a flat target space and its generalizations to curved backgrounds for specific values of $N$ in the next sections.
	
	\subsection{$N=1$ and $U(1)$ background}
	\label{sub1}
	
	When dealing with $1$ odd coordinate and the induced supersymmetry transformation of the worldline, a ghost degree $1$ element that behaves as a differential is 
	\begin{equation}
		Q = c H + \gamma q_0 - \gamma^2 b,
	\end{equation} 
	for $H := \frac12 P^2$ and $q_0 := \Psi^\mu P_\mu$ as before. BRST cohomology yields equivalent results to the quantization covered in the previous section: there is a Dirac spinor in the cohomology group $H^0(Q, \mathcal{H})$ at ghost degree $0$, for $\mathcal{H}$ a Hilbert space containing \eqref{Dirac} as well as some ghost states. 
	
	Let us now instead consider a coupling to the electromagnetic potential:
	\begin{equation}
		\delta q = \Psi^\mu A_\mu(x), \qquad A_\mu(x) \mathrm{d}x^\mu \in \Omega^1(M, \mathrm{u}(1))
	\end{equation}
	A first condition to ensure nilpotency, if by $q$ we denote $q := q_0+\delta q $ and by $\Pi_\mu$ the covariant derivative $\Pi_\mu := P_\mu + A_\mu$, is that $H$ must be:
	\begin{equation}
		H = \{q,q\} = \{\Psi^\nu \Pi_\nu, \Psi^\rho \Pi_\rho\} = \{\Psi^\nu,\Psi^\rho\}\Pi_\rho \Pi_\nu + \Psi^\nu\Psi^\rho[\Pi_\nu,\Pi_\rho]. 
	\end{equation}
	Then a second condition comes from $[q, H] \overset{!}{=}0$, whose explicit calculation can be instructive. Using just Leibniz rule, as well as symmetry/antisymmetry arguments, we can suggest the following algebraic manipulations:
	\begin{align}
		[q, H] = & \, \Psi^\mu\Psi^\nu\Psi^\rho [\Pi_\mu,[\Pi_\nu,\Pi_\rho]] + \{\Psi^\mu,\Psi^\nu\}\Psi^\rho [\Pi_\nu,\Pi_\rho]\Pi_\mu - \Psi^\nu \{\Psi^\mu, \Psi^\rho\}[\Pi_\nu, \Pi_\rho]\Pi_\mu \notag \\
		\, & + \Psi^\mu \{\Psi^\nu,\Psi^\rho\} \left([\Pi_\mu,\Pi_\rho]\Pi_\nu + \Pi_\rho[\Pi_\mu, \Pi_\nu]\right) \\
		= & \, \Psi^\mu\Psi^\nu\Psi^\rho [\Pi_\mu,[\Pi_\nu,\Pi_\rho]] + \{\Psi^\mu,\Psi^\nu\}\Psi^\rho [\Pi_\mu,[\Pi_\rho, \Pi_\nu]] \\
		= & \, \left(\Psi^\mu\Psi^\nu\Psi^\rho \pm \Psi^\nu \Psi^\mu\Psi^\rho \right)  [\Pi_\mu,[\Pi_\nu,\Pi_\rho]] + \Psi^\rho [\Pi_\mu, [\Pi_\rho,\Pi^\mu]] \\
		= & \Psi^{[\mu} \Psi^{\nu]} \Psi^\rho [\Pi_\mu, [\Pi_\nu, \Pi_\rho]]
	\end{align}
	The conclusion follows from Jacobi identity (Bianchi identity for the covariant derivative $\Pi$): The new operator is nilpotent regardless of whether or not $A_\mu$ is a Maxwell field, i.e. if it fulfills the Maxwell equations for the $U(1)$-gauge theory.
	
	\subsection{$N=2$ and $SU(n)$ background}
	
	A $2$-supersymmetric worldline is associated with a spin-$1$ field, which is a gauge boson for either an abelian gauge theory or Yang-Mills gauge theory \cite{Dai:2008bh}. Since in the present situation there are two copies of $\Psi^\mu$, it is convenient to form complex linear combinations and therefore represent the fundamental and antifundamental representation of $SU(2)$ by the homomorphisms $\text{Hom}(SU(2), \text{End}(\mathbb{C}^2))$:
	\begin{equation}
		\Psi^\mu = \frac{1}{\sqrt{2}} (\Psi^\mu_1 + \mathrm{i}\Psi^\mu_2), \qquad \bar{\Psi}^\mu = \frac{1}{\sqrt{2}} (\Psi^\mu_1 - \mathrm{i}\Psi^\mu_2),
	\end{equation}
	so that the anticommutators lead to:
	\begin{equation}
		\{\Psi^\mu, \bar{\Psi}^\nu\} = \eta^{\mu\nu} = \{\bar{\Psi}^\mu, \Psi^\nu\}, \qquad \{\bar{\Psi}^\mu, \bar{\Psi}^\nu\} = 0 = \{\Psi^\mu, \Psi^\nu\}.
	\end{equation}
	Notice that there is an extra $SO(2) \equiv U(1)$ group that acts on the space spanned by the Gamma matrices $\Psi$ and $\bar\Psi$, and it is easy to be convinced that a good guess for the $U(1)$ generator is $J = \Psi^\mu\bar{\Psi}_\mu - 1$:
	\[
	[J, \Psi^\nu] = 2 \Psi^\nu, \qquad [J, \bar{\Psi}^\nu] = -2 \bar{\Psi}^\nu.
	\]
	
	\textbf{BRST cohomology.} In the previous \cref{sub1} we did not analyse the ghost degree $0$ cohomology of $Q$ because it coincides with Dirac quantization very straightforwardly. For the current case let us unveil the cocycles and coboundaries of the BRST differential. At ghost degree zero, the cocycles are the field equations, instead the coboundaries are the gauge symmetries. First of all, we need a Hilbert space where the differential will act. The Hilbert space is constructed as the representation space of a maximal commuting subalgebra of the BRST algebra\footnote{With obvious reference to $\Psi$ and $\bar\Psi$, $(\gamma,\bar\gamma, \beta,\bar\beta)$ are the complex linear combinations such as: $\gamma = \frac{1}{2} (\gamma_1 + \mathrm{i}\gamma_2)$ and $\bar\gamma = \frac{1}{2} (\gamma_1 - \mathrm{i}\gamma_2)$}
	\[
	[X^\mu,P_\nu] = \mathrm{i}\delta^\mu_\nu, \quad \{\Psi^\mu, \bar{\Psi}^\nu\} = \eta^{\mu\nu}, \quad [\bar\gamma, \beta] = 1 = [\gamma, \bar\beta], \quad \{c,b\}=1.
	\] 
	For example the set $(X^\mu,\Psi^\mu,\gamma, \beta, c)$ can be taken as creation operators, while all the barred operators will annihilate the highest weight vector $\ket{0} \in \mathcal{H}$. 
	To limit the number of states from above, it is convenient to restrict to $\ker J$ in $\mathcal{H}$, where the "number" constraint yields the BRST operator $J$ given by:
	\begin{equation}
		J = \Psi \cdot \bar{\Psi} +  \beta \bar{\gamma} - \gamma \bar\beta -1. \label{J}
	\end{equation}
	The space $\ker J$ in $\mathcal{H}$ physically correspond to the vector space of $1$-particles. Hence a generic state in $ \mathcal{H} \cap \ker J$ with $C^\infty(M)$ coefficients is:
	\begin{equation}
		\left(A_\mu(x) \Psi^\mu + \gamma f(x) + \beta \tilde g(x) + c \big(\tilde A_\mu(x) \Psi^\mu + \gamma \tilde f(x) + \beta g(x)  \big)\right)\ket{0} .
	\end{equation} 
	A cocycle or otherwise the $\ker Q$ for flat derivatives, i.e. $Q = cP^2 + \bar\gamma \Psi\cdot P + \gamma \bar{\Psi}\cdot P - \gamma\bar\gamma b$, is the following: 
	\begin{align}
		c P^2 A_\mu(x) \Psi^\mu \ket{0} - c \Psi \cdot P g(x) \ket{0} = 0, \label{YM1}\\
		\gamma P \cdot A(x) \ket{0} - \gamma g(x) \ket{0} = 0. \label{YM2}
	\end{align}
	A further equation for the field $f(x)$ is omitted. Replacing $g$ from \eqref{YM2} into \eqref{YM1} one finds the linearized Yang-Mills equations, at first order in $A$.
	
	The coboundaries in ghost degree $0$ stems from the ghost degree $-1$ state $\ket{\lambda} = \lambda(x)\beta \ket{0}$:
	\begin{equation}
		\delta A_\mu(x) \Psi^\mu \ket{0} = \{Q, \ket{\lambda}\} = \Psi^\mu P_\mu \lambda(x) \ket{0},
	\end{equation}
	which is the sought-after $U(1)$ gauge symmetry.
	
	\textbf{Curved $SU(n)$ background.} In the $N=2$ case with "covariant" charges $q = \Psi^\mu \Pi_\mu$ and $\bar{q} = \bar{\Psi}^\nu \Pi_\nu$,\footnote{The non-abelian indices are omitted.} the BRST differential is
	\begin{equation}
		Q = cH + \gamma \bar q + \bar\gamma q - \gamma \bar\gamma b,
	\end{equation}
	and therefore
	\begin{align}
		\{\gamma \bar q,\gamma \bar q\} + 2\{\bar \gamma q , \gamma \bar q\} + \{\bar\gamma q, \bar\gamma q\} =& \, 2 \gamma \bar\gamma \{\Psi^\nu, \bar{\Psi}^\rho\} \Pi_\rho \Pi_\nu + \big( \bar\gamma^2\Psi^\nu\Psi^\rho + \gamma\bar\gamma \bar\Psi^\nu\Psi^\rho \notag\\
		& + \gamma^2\bar\Psi^\nu \bar\Psi^\rho + \gamma\bar\gamma \Psi^\nu\bar\Psi^\rho \big) [\Pi_\nu,\Pi_\rho]. \label{qq2}
	\end{align}
	Compared to the previous subsection where we dealt with the $N=1$ particle, triplets of $\Psi$ and $\bar{\Psi}$ will arise when proceeding further with the nilpotency check. As long as they are of the same type, we could use Jacobi identity, but for mixed terms this is not sufficient. The clever observation of \cite{Dai:2008bh} was to evaluate the expression on the vectors lying in $\mathcal{H}$ and in the kernel (in $\mathcal{H}$) of the BRST "number" operator $J$ \eqref{J}, since effectively, any state in $\mathcal{H} \cap \ker J$ consists of $1$-particles and therefore will be annihilated by the action of $n \geq 2$ barred operators. Thus \eqref{qq2} boils down to:
	\begin{equation}
		\gamma \bar\gamma \{q, \bar q\}\vert_{\mathcal{H}\cap \ker J} = \gamma \bar\gamma \big(2\{\Psi^\nu, \bar{\Psi}^\rho\}\Pi_\rho \Pi_\nu  + \alpha (\Psi^\nu\bar\Psi^\rho - \Psi^\rho\bar\Psi^\nu) [\Pi_\nu,\Pi_\rho] \big)\vert_{\mathcal{H}\cap \ker J}.
	\end{equation}
	We have the freedom to choose a $\mathbb{C}$-number $\alpha$ for the second term because that is in fact made up with two annihilators (barred operators), but $\alpha\neq 0$ for later convenience. From $[\gamma\bar q +\bar\gamma q, \{q,\bar q\}]$ we hence find 
	\begin{align}
		\frac{1}{2}[H,\bar\gamma q +\gamma \bar q]= & (\gamma \bar\Psi^\mu + \bar\gamma \Psi^\mu) [\Pi_\mu, \Pi^2] + \alpha(\gamma \bar\Psi^\mu +\bar\gamma \Psi^\mu)\Psi^{[\nu} \bar\Psi^{\rho]} \Pi_\mu [\Pi_\nu,\Pi_\rho] \notag \\
		\, & - \alpha(\Psi^{[\nu} \bar\Psi^{\rho]})(\gamma\bar\Psi^\mu + \bar\gamma \Psi^\mu) [\Pi_\nu,\Pi_\rho]\Pi_\mu \\
		= &  (\gamma \bar\Psi^\mu + \bar\gamma \Psi^\mu) [\Pi_\mu, \Pi^2] + \alpha(\gamma \bar\Psi^\mu +\bar\gamma \Psi^\mu)\Psi^{[\nu} \bar\Psi^{\rho]} (\Pi_\mu[\Pi_\nu,\Pi_\rho] -[\Pi_\nu,\Pi_\rho]\Pi_\mu) \notag \\
		& + \alpha(\gamma \eta^{\mu[\nu} \bar\Psi^{\rho]} - \bar\gamma\Psi^{[\nu} \eta^{\rho]\mu})[\Pi_\nu,\Pi_\rho]\Pi_\mu
	\end{align}
	Setting $\alpha=2$, moving all annihilators to the right and forgetting the terms that kill the restricted Hilbert space, we get:
	\begin{align}
		\frac{1}{2}[H,\bar\gamma q +\gamma \bar q] =&  (\gamma \bar\Psi^\mu + \bar\gamma \Psi^\mu) [\Pi^\rho, [\Pi_\mu, \Pi_\rho]] + 2 \gamma \bar\Psi^\mu [\Pi^\rho, [\Pi_\rho,\Pi_\mu]]
	\end{align}
	Hence we recover the result that the deformed BRST operator $Q$ is nilpotent iff Yang-Mills equations are fulfilled:
	\begin{equation}
		\nabla \star F = 0, \qquad \nabla:= -\mathrm{i}\Pi, \; F \in \Omega^2, \, F := [\nabla_\mu, \nabla_\nu] \mathrm{d}x^\mu \wedge \mathrm{d} x^\nu ,
	\end{equation}
	where $\star$ is the Hodge star operator. This observation is quite remarkable: compared to the previous \cref{sub1} the target space seen by the particle is now dynamical.
	
	\subsection{$N=4$ and gravity backgrounds}
	
	In the $N=4$ case there are first-quantized spin-2 particles in the first BRST cohomology group. The larger amount of parity odd coordinates makes the formulas more involved. Hence we will be sketchy and just briefly enunciate the results of \cite{Bonezzi:2018box} and \cite{Bonezzi:2020jjq}. Now the complex linear combinations of Gamma matrices enjoy an $SO(4)$ symmetry, which can be used to restrict the Hilbert space. In \cite{Bonezzi:2018box} the full $\mathfrak{so}(4)$ algebra was used, whereas in \cite{Bonezzi:2020jjq} the authors resorted to the subgroup $U(1)\times U(1) \subset SO(4)$. Let us point out that in ghost degree $0$, the cohomology of the differential $Q + \tau^i J_i$ on the Hilbert space coincides with the cohomology of $Q$ on $\mathcal{H} \overset{i}{\cap} \ker J_i$ when the $J_i$ are charges for the $U(1)\times U(1)$ subgroup (hence $i=1,2$). For the more general case this is not necessarily true and truncating to $\ker J_i$ is a very strong requirement.
	
	To summarize, it was respectively proven that:
	\begin{itemize}
		\item for a maximal gauge fixing of $SO(4)$ the reduced Hilbert space contains only the graviton, auxiliary fields and ghost fields. Acting with $n\geq 3$ barred operators annihilate every state in the vector space. Eliminating terms with such a number of barred operators in the calculation of $Q^2$, with covariant $Q$ constructed with the Levi-Civita connection, the obstruction to be zero is exactly Einstein's equations;
		\item when gauge fixing $U(1)\times U(1)$, then the Hilbert space restricted to the kernel of the number operator for $U(1)\times U(1)$ contains also a $2$-form $B$ and a dilaton. 
		Nilpotency of a $Q$ constructed with a non-symmetric covariant derivative (which has $H = \mathrm{d} B$) is ensured iff the $(g,H,\phi)$ triplet satisfies the Supergravity equations of motion. The observation that the vector space of states is (contained) in the kernel of polynomials with a number $n\geq 3$ of barred operators, is again used to prove the result. 
	\end{itemize}

	\section{Conclusion}
	\label{4}
	
	The cohomological methods of BRST and BV have proven useful in perturbative quantization, when attempting to calculate the path integral of a degenerate action, i.e. an action that enjoys some gauge symmetries. Besides keeping covariance explicit, BRST (and BV) allows to consider more general, curved backgrounds in target space. In this short note we have reviewed the case of  $N=1,2,4$ supersymmetry of particles in the worldline, and assumed interactions with a gauge potential due to bosons of the same spin (or higher, in the $N=1$ case) as that of the first-quantized particles found in the BRST-cohomology. Interactions between gauge bosons and physical states of the Hilbert space were implemented semiclassically through a deformed BRST differential. We saw that the dynamics of the background field was not fixed in the $N=1$ case, whereas for $N=2$ the Yang-Mills background was found to satisfy the Yang-Mills equations. Eventually $N=4$, admitting a spin-2 field, had a a deformed $Q$ which behaved as a differential iff Einstein's equations or $(g,H,\phi)$ Supergravity equations were satisfied. A crucial step in the proofs was to restrict the Hilbert space to a certain fixed value of multiparticle states ($1$-particles for Yang-Mills, $2$-particles for (Super-)gravity). This truncation of the Hilbert is actually more than just a trick to implement nilpotency of a generic operator: the BRST operator $Q + \tau^iJ_i$ (extended with the number operator for $SO(N)$ symmetry) should yield non-trivial $Q$-cohomology only when looking at $H^\bullet(Q,\ker J\vert_\mathcal{H})$. This happens because the number operator $J$ has a simple diagonal multiplicative action and therefore any state that is not in its kernel is always $J$-exact.

	An important step towards the recovery of the full Supergravity equations would be to study Ramond-Ramond forms in target spaces. However, in superstrings, these $p$-forms come from target space spinors defined on the worldsheet. Then the tensor product decomposition of two spinors decomposes into the direct sum of $p$-forms (the Ramond-Ramond forms). By analogy with the string, in the superworldline for a relativistic massless particle we are thus forced to introduce further fields with spinorial indices. An available option would be to "resolve" the Gamma matrix $\Psi^\mu$, in the following fashion:
	\begin{equation}
		\Psi^\mu = \theta^\alpha \sigma^\mu_{\alpha \dot\beta} \tilde\lambda^\beta + \tilde\theta_{\dot\alpha} \tilde\sigma^{\mu \, \dot\alpha \beta} \lambda_\beta,
	\end{equation}
	and only later check whether $\Psi^\mu$ still satisfies the Clifford algebra. The spinors $\theta$ and $\lambda$ and their chiral companions are also required to have opposite Grassmann degree (resp. odd and even), and they respect some relations. Moreover $\sigma^\mu = (\mathbb{1}, \sigma^i)$ are the extended Pauli matrices and $\tilde\sigma^{\mu \, \dot\alpha\beta} = \epsilon^{\dot\alpha \dot\gamma}\epsilon^{\beta\gamma}\sigma^\mu_{\gamma\dot\gamma}$. The outcomes of this ansatz are discussed in a separate article \cite{Boffo:2022pbs}.
	
	\bigskip
	
	\textbf{Acknowledgements.} A warm thank to the organizers of the 42th Srn\'{i} Winter School Geometry and Physics 15-22 January 2022. I am glad that I had the chance to participate and present my research there. A big thank you to Ivo Sachs for introducing me to the topic and explaining to me many aspects of it. I acknowledge the GA\v{C}R grant EXPRO 19-28628X for financial support. 
	
	\bibliographystyle{ieeetr}
	\bibliography{res-stat}
\end{document}